\documentclass[12pt,preprint]{aastex}

\shorttitle{Dipolar Evolution in a Coronal Hole Region}
\shortauthors{Yang et al.}

\begin{document}

\title{Dipolar Evolution in a Coronal Hole Region}

\author{Shuhong Yang\altaffilmark{}, Jun Zhang\altaffilmark{}}

\affil{Key Laboratory of Solar Activity, National Astronomical
Observatories, Chinese Academy of Sciences, Beijing 100012, China}
\email{[shuhongyang;zjun]@ourstar.bao.ac.cn}

\and

\author{Juan Manuel Borrero\altaffilmark{}}
\affil{Max Planck Institut for Solar System Research, Max Planck
Strasse 2, 37191, Katlenburg-Lindau, Germany}
\email{borrero@mps.mpg.de}


\begin{abstract}

Using observations from the SOHO, STEREO and Hinode, we
investigate magnetic field evolution in an equatorial coronal hole
region. Two dipoles emerge one by one. The negative element of the
first dipole disappears due to the interaction with the positive
element of the second dipole. During this process, a jet and a
plasma eruption are observed. The opposite polarities of the
second dipole separate at first, and then cancel with each other,
which is first reported in a coronal hole. With the reduction of
unsigned magnetic flux of the second dipole from
9.8$\times$10$^{20}$ Mx to 3.0$\times$10$^{20}$ Mx in two days,
171 {\AA} brightness decreases by 75\% and coronal loops shrink
obviously. At the cancellation sites, the transverse fields are
strong and point directly from the positive elements to the
negative ones, meanwhile Doppler red-shifts with an average
velocity of 0.9 km s$^{-1}$ are observed, comparable to the
horizontal velocity (1.0 km s$^{-1}$) derived from the cancelling
island motion. Several days later, the northeastern part of the
coronal hole, where the dipoles are located, appears as a quiet
region. These observations support the idea that the interaction
between the two dipoles is caused by flux reconnection, while the
cancellation between the opposite polarities of the second dipole
is due to the submergence of original loops. These results will
help us to understand coronal hole evolution.

\end{abstract}


\section{Introduction}

Coronal holes (CHs) are dark and void areas on the Sun if observed
with X-ray (Underwood \& Muney 1967) and EUV lines (Reeves \&
Parkinson 1970). They are low density and temperature regions
compared with the quiet Sun (QS) (Munro \& Withbroe 1972; Harvey
1996; Chiuderi Drago et al. 1999). In CHs, magnetic fields are
dominated by one polarity and open magnetic lines are concentrated
(Bohlin 1977). These open magnetic lines extend to the
interplanetary space along which plasma escapes, giving rise to
fast solar wind (Krieger et al. 1973; Zirker 1977; Wang et al.
1996; Harvey \& Recely 2002; Tu et al. 2005). However, magnetic
fields are not exclusively unipolar in CHs and there also exist
closed coronal loops besides the open flux (eg. Levine 1977; Zhang
et al. 2006). Wiegelmann \& Solanki (2004) computed some
properties of coronal loops in CHs and the QS. They found that
high and long closed loops in CHs are extremely rare, whereas
short and low-lying loops are almost as abundant as in the QS.
Fisk (2005) predicted that a CH is a region with a local minimum
in the rate of emerging dipoles. This prediction was supported by
Zhang et al. (2006), whose results reveal that the dipole
emergence rate in the QS is 4.3 times as large as that in the CH.
According to Fisk \& Schwadron (2001) and Fisk (2005), dipoles can
transport open flux on the solar surface through magnetic flux
reconnection and result in the CH evolution by affecting magnetic
field distribution and characters.

Magnetic flux cancellation is an observational phenomenon of
magnetic flux disappearance in the encounter of two magnetic
elements of opposite polarities (Martin et al. 1985; Livi et al.
1985; Zhang et al. 2001). Zwaan (1978, 1987) illustrated three
modes for removal of magnetic flux with different polarities from
the photosphere: (1) If two poles are still connected by coronal
loops, then the disappearance of the magnetic flux can be caused
by retraction of flux loops to below the photosphere; (2) When two
poles with no initial connection encounter, magnetic reconnection,
creating both $\Omega$$-$shaped and U$-$shaped loops, is required
to remove the magnetic flux from the photosphere. If reconnection
occurs below the photosphere, the newly formed U$-$shaped loop is
pulled out of the photosphere; (3) If reconnection takes place
above the photosphere, the $\Omega$$-$shaped loop submerges
(retracts) below the photosphere. Magnetic reconnection may not be
necessary for forming the emerging U$-$shaped loop if two poles
connect below the photosphere (Parker 1984; Lites et al. 1995).

Generally, dipoles appear to rise from below the solar surface,
separate and dissipate (Wallenhorst \& Topka 1982; Liggett \&
Zirin 1983). However, some dipoles do not decouple from subsurface
fields (Zirin 1985). An excellent example exhibiting the
submergence of part of an active region was reported by Rabin et
al. (1984). Another example displaying the submergence of entire
region is the disappearance of a sunspot group, studied by Zirin
(1985) with videomagnetograms and H$\alpha$ observations.

When collision occurs between opposite polarities, transverse
fields and Doppler velocities at the cancellation sites have been
studied by many authors. Zhang et al. (2009) found that transverse
fields connecting cancelling magnetic elements are formed. Harvey
et al. (1999) estimated vertical velocity of magnetic flux descent
ranging from about 0.3 to 1.0 km s$^{-1}$, and Yurchyshyn \& Wang
(2001) observed an upflow of about 0.6 km s$^{-1}$ at a
cancellation site. The first direct observational evidence of flux
retraction in cancelling magnetic features was presented by Chae
et al. (2004). They found that the magnetic fields were nearly
horizontal at the place where two opposite polarities cancelled
each other. In addition, they observed significant magnetic flux
submergence of about 1.0 km s$^{-1}$ near the polarity inversion
line. Recently, magnetic filed properties at the cancellation
sites were investigated by Kubo \& Shimizu (2007) based on more
collision events.

Due to the restriction of observations, not all the properties of
CHs are well known, such as CH formation and decay. These CH
properties are related to magnetic field structures and evolution,
the key characters to understand most solar phenomena, so it is
still important for us to investigate magnetic fields in CHs with
high spatial and temporal resolution data, especially the vector
fields and other plentiful information from the Hinode (Kosugi et
al. 2007). In this work, we study the evolution process of two
dipoles in an equatorial CH, using observations from the Solar and
Heliospheric Observatory (SOHO; Domingo et al. 1995), the Solar
Terrestrial Relations Observatory (STEREO; Howard et al. 2008;
Kaiser et al. 2008) and the Hinode. We describe the observations
in Sect. 2 and data analysis in Sect. 3. The two parts of Sect. 4
show respectively the interaction between the two dipoles and the
cancellation between the opposite polarities of the second dipole.
The conclusions and discussion are presented in Sect. 5.

\section{Observations}

The CH shown in Fig. 1 was observed with the Michelson Doppler
Imager (MDI; Scherrer et al. 1995) and the Extreme-ultraviolet
Imaging Telescope (EIT; Delaboudini\`{e}re et al. 1995) on SOHO,
the Extreme Ultra Violet Imager (EUVI; Howard et al. 2008), part
of the Sun-Earth Connection Coronal and Heliospheric Investigation
(SECCHI; Howard et al. 2008) aboard STEREO, combined with the
Spectro-Polarimeter (SP; Lites et al. 2001) and the Narrowband
Filter Imager (NFI; Kosugi et al. 2007), two components of the
Solar Optical Telescope (SOT; Ichimoto et al. 2008; Shimizu et al.
2008; Suematsu et al. 2008; Tsuneta et al. 2008) on board Hinode.

We adopt the MDI, EIT and EUVI full-Sun observations between 08:00
UT 02 March and 12:00 UT 07 March 2008. The data sets used in this
study are summarized in Table 1. MDI provides full-disk
longitudinal magnetograms with pixel sampling of 1{\arcsec}.97.
The cadence of MDI magnetograms was 1 minute during the periods of
11:35 UT -- 14:36 UT 02 March and 00:57 UT 03 March -- 12:00 UT 07
March, and 96 minutes during the other periods. EIT observed the
CH with a pixel size of 2{\arcsec}.63 and a cadence of 12 minutes
in Fe XII 195 {\AA}. STEREO A and B simultaneously observed
full-Sun with a 1{\arcsec}.59 pixel resolution and a 2.5-min
cadence in Fe IX 171 {\AA}. On the early of 05 March 2008, when
the CH was located at disk center viewed from the Earth, the
STEREO A and B were 44$\degr$ ahead and 48$\degr$ behind the
Earth. In order to maintain the CH close to disk center in each
image, we employ STEREO B observations before 00:00 UT 05 March,
and STEREO A observations after that.

Hinode SP and NFI observed only part of the CH, but covered
majority of our target during 03 March and 05 March, 2008. The SP,
which provides full Stokes profiles (I, Q, U and V) of Fe 6301.5
{\AA} ($g_{eff}=1.67$) and 6302.5 {\AA} ($g_{eff}=2.5$) lines,
scanned the target along East$-$West direction in the fast map
mode with a step of 0{\arcsec}.295, and the pixel size along the
slit is 0{\arcsec}.32. The NFI obtained the Stokes I and V images
at an offset wavelength of $-$172 m{\AA} from the center of the Na
5896 {\AA} line with a cadence of 3 minutes and a pixel resolution
of 0{\arcsec}.16.

\section{Data analysis}

At first, all the images from the EIT, EUVI and NFI are prepared
by applying the standard processing routines, including flat field
correction, dark current and pedestal subtraction, cosmic ray
removal, et al.. Then we rotate the target in the MDI, EIT and
EUVI images to the central meridian. After that we re-scale and
uniform their pixel sizes. In order to remove the drift due to the
correlation tracker motion (Shimizu et al. 2008), NFI V/I (i.e.
Stokes V signals divided by Stokes I signals obtained at the same
time) images are coaligned with each other by using the image
cross-correlation method. Then the MDI magnetograms were aligned
to the NFI V/I images by cross-correlating specific features after
being re-scaled. The same coalignment method is also applied to
match the EIT and EUVI images. Since the MDI and the closest EIT
images are from the same satellite and can be coalined easily, we
think the MDI, EIT, EUVI and NFI images are coaligned well now.
Due to the high conductivity, the coronal plasma is frozen into
the magnetic field. Consequently, the emitting plasma structures
outline the magnetic field lines (Wiegelmann \& Solanki 2004). So
in order to identify loop connections well (see Kano \& Tsuneta
1995; Kubo \& Shimizu 2007), the EUVI structures are contrast
enhanced by using an unsharp mask filter (refer to Feng et al.
2007).

The spectropolarimetric data from Hinode/SP are processed with the
routine \emph{sp$_{-}$prep.pro} available in the \emph{Solar
Software (SSW)} package. This routine performs standard
calibrations such as flat-fielding, dark current correction,
polarimetric and instrumental calibration (Ichimoto et al. 2008;
Kubo et al. 2008). The calibrated Stokes profiles are analyzed
using the VFISV inversion code (Borrero et al. 2009). This code
uses the Milne$-$Eddington solution for the radiative transfer
equation to produce synthetic Stokes profiles that are then
compared with the observed ones. The free parameters of the model
are: magnetic field strength $B$, inclination ($\gamma$) and
azimuth ($\phi$) of the magnetic field vector in the observer's
reference frame, line-of-sight (LOS) velocity $V_{los}$, continuum
to core absorption coefficient $\eta_0$, Doppler width $\Delta
\lambda_D$, source function and source function gradient $S_0$ and
$S_1$, and finally the filling factor of the magnetic component
$\alpha_{mag}$. We do not consider the damping parameter and
macroturbulent velocities as free parameters since their effect
can be mimicked by the other thermodynamic parameters and they do
not affect the determination of the important quantities such as
the magnetic field vector and velocity. With this, we have a total
number of 9 free parameters, which are iteratively modified (using
the Levenberg$-$Marquardt non-linear least squares fitting
technique) in other to achieve a better match between synthetic
and observed profiles.

The non-magnetic component is obtained by averaging the pixels of
the map that possess a polarization signal below the noise level
(1.2$\times$10$^{-3}$\emph{I$_{c}$}, where \emph{I$_{c}$} is the
continuum intensity). The same non-magnetic component is used in
the inversion of all pixels in the map. Our approach is therefore
slightly different from that of Orozco Su\'arez et al. (2007), who
employed a local (average around each inverted pixel) non-magnetic
component. We have also repeated our inversions following that
approach, but our results did not change. Given the small amount
of scattered light known to exist within the SP instrument
on-board Hinode (Danilovic et al. 2008), the amount of
non-magnetic component $(1-\alpha_{mag})$ can be either
interpreted as true non-magnetic unresolved component inside the
SP resolution element or as a degradation of the polarization
signal due to diffraction (Orozco Su\'arez et al. 2007).

The zero LOS velocity $V_{los}$ has been obtained by calculating
the convective blue-shift in the Fe I 6302.5 \AA~ line using the
Fourier Transform Spectrometer atlas at disk center (FTS; Brault
\& Neckel 1987) and also by subtracting the solar gravitational
red-shift. Since the observed region was located close to disk
center no further corrections due to solar rotation were
necessary.

In the vector field measurements based on the Zeeman effect, there
exists a 180{\degr} ambiguity in determining the field azimuth,
$\phi$. However, it is not unresolvable. Potential field
approximation is one of the fairly acceptable methods to resolve
the ambiguity (Lites et al. 1995). We construct the photospheric
vector magnetic fields by computing the potential fields using the
SP longitudinal megnetogram, and the azimuth angles with
180{\degr} ambiguity are disambiguated by being selected and
determined to be close to that we have constructed. Finally, the
SP maps are coaligned with the NFI images by using the SP
longitudinal magnetogram.

By using the similar method introduced in Chae et al. (2007), we
convert the Na V/I signals to the LOS magnetic fields according to
the temporally closest SP data. A linear relation,
$B_{los}$=$\beta$$\times$V/I, between the circular polarization,
V/I, and the line-of-sight field strength, $B_{los}$, is applied.
We determine the calibration coefficient, $\beta$, for each
interval partitioned by V/I=$-$0.07, $-$0.06, ..., 0.09. The 18
values of $\beta$ range from 4.0 kG to 12.3 kG and the mean
calibration coefficient is 6.8 kG.

The CH boundary (see Fig. 1) is determined with the brightness
gradient method which was developed by Shen et al. (2006; see also
Luo et al. 2008). In an EUVI 284 {\AA} image, the pixel value, b,
varies in a range. For any value of b, we can plot a contour and
calculate the area, A, enclosed by it. The CH boundary is at the
place where
\emph{f}=$\delta$\emph{b}$/$$\delta$\emph{A}=\emph{f}$_{max}$.

\section{Results}

The CH in this study is dominated by the positive polarity (see
the bottom left panel in Fig. 1). Within a
90{\arcsec}$\times$90{\arcsec} region (outlined by white squares
in Fig. 1), two dipoles emerge one by one. We focus on the
interaction between the two dipoles and the disappearance of the
second dipole from 02 March to 07 March, 2008.

\subsection{Interaction between the two dipoles}

Figure 2 presents the interaction process between the two dipoles.
Top panels are time sequence of MDI magnetograms, which show the
emergence and interaction of the two dipoles. The middle panels
exhibit the coronal response in EUVI 171 {\AA} line. Loop
connections can be seen much more clearly in the contrast enhanced
171 {\AA} images (bottom panels). The positive element ``A" of the
first dipole (denoted by arrows ``1") appeared obvious at 09:39 UT
02 March and then grew larger, and brightening point appeared
simultaneously at the corresponding location as shown in the
leftmost column. Then the negative element ``B" began to emerge
and the dipole ``1" reached its maximum size with total unsigned
LOS flux of 2.0$\times10$$^{20}$ Mx at 20:51 UT 02 March. At this
time, the loop connections between elements ``A" and ``B" are
emphasized with green curves in the contrast enhanced image at
20:47 UT 02 March. The second dipole (indicated by arrows ``2")
started to appear in the magnetogram from 20:51 UT 02 March with
slightly distinguishable features and became quite obvious at
22:27 UT 02 march. The magnetogram acquired at 00:03 UT on 03
March shows that its positive element ``C" was located in contact
with element ``B". Element ``B" split into two segments during its
disappearance process, as shown in the magnetogram at 03:11 UT 03
March. Both two segments of ``B" disappeared completely at 08:22
UT 03 March. Element ``A" moved toward ``C" and merged with ``C"
into ``A$+$C", still called element ``C" by us considering the
small size of ``A" compared to that of ``C". Dipole ``2" continued
emerging and reached its maximum with total unsigned LOS flux of
9.8$\times10$$^{20}$ Mx at 19:01 UT 03 March (top right panel).

During the interaction process between the two dipoles, a jet and
a plasma eruption have been observed, as shown in Fig. 3. At the
early emerging stage of the dipole ``2", there was a small bright
point (denoted by arrow ``1") at the adjacent area of elements
``B" and ``C" (top left panel). Two and a half minutes later, a
jet (indicated by arrows ``2") rooting at the bright point was
observed. The lifetime of the jet is only 5 minutes. The second
dipole continued to emerge and another obvious brightening point
(indicated by arrow ``3") appeared at the contacted region of ``B"
and ``C" (bottom left panel). At 23:46 UT, a cloud of plasma was
disturbed (denoted by arrow ``a") and erupted five minutes later
(arrow ``b"). After the eruption, a dimming area appeared (arrow
``c").

\subsection{Magnetic flux cancellation of the second dipole}

Two elements of dipole ``2" separated as flux continually emerged
until 19:01 UT 03 March. Then they began to cancel with each
other, as shown in Fig. 4. Top five panels display the evolution
of dipole ``2" in five days. When the dipole well developed, the
flux of each polarity was almost concentrated. As the cancellation
began, each pole broke into several fragments. Then these
fragments with opposite polarities moved together and cancelled
gradually. Accompanying the cancellation process, the
corresponding coronal region in 171 {\AA} images became darker
(middle panels), while the general loop connections were not
changed much (bottom panels). However, the loop systems became
fewer and fuzzier, and the length of loops became shorter. At
20:47 UT 03 March, the amount of identifiable loops was 7 while
there was no more than 4 at 06:23 UT 05 March. At 19:01 UT 03
March, when the dipole well developed, the length of the longest
loop was about 30 Mm. Two days later, most of the long coronal
loops were about 15 Mm. At 06:24 07 March, only small dispersed
elements of dipole ``2" remained (bottom right panel in Fig. 1).
The northeastern part of the CH, where the dipoles were located,
appeared as a quiet region and the underlying magnetic fields
evolved to mixed polarities (see the rectangle region in Fig. 1).

Figure 5 shows the temporal variations of negative and positive
magnetic flux in MDI magnetograms and brightness in 171 {\AA}
images obtained from the ellipse region in Fig. 4. In two days,
both the negative and positive flux reduced smoothly by
3.4$\times$10$^{20}$ Mx at a steady cancellation rate of
0.7$\times$10$^{19}$ Mx h$^{-1}$. During this period, the
brightness in 171 {\AA} images decreased by about 75\%.

The Na V/I high spatial resolution magnetograms are used to study
the cancellation between two polarities of dipole ``2" in details.
Two obvious cancellation processes are displayed in Fig. 6. In the
circle region (top panels), a negative flux island moved
straightly toward the positive flux, cancelled  with it and
disappeared totally at 14:29 UT 04 March. In the octagon area
(bottom panels), another flux island observed on 05 March also
cancelled with the positive flux. We measure the cancellation
rates between the opposite polarities of the second dipole in the
calibrated Na V/I magnetograms (within the rectangle area outlined
in Fig. 4). It indicates that both the positive and negative
polarities decreased at an average rate of 0.8$\times$10$^{19}$ Mx
h$^{-1}$ during the period of 11:20--16:17 on 04 March and of
0.7$\times$10$^{19}$ Mx h$^{-1}$ from 11:39 to 15:27 on 05 March.
We also measure the disappearing rate of the two small negative
islands shown in Fig. 6. At the pre-cancellation stage, the flux
of the two islands is $-$0.4$\times$10$^{19}$ Mx (outlined with
circles in the top panels) and $-$0.5$\times$10$^{19}$ Mx
(outlined with octagons in the bottom panels), respectively. Both
the two islands disappeared at a mean rate of 0.3$\times$10$^{19}$
Mx h$^{-1}$.

The negative island shown in the circle region in Fig. 6 moved
toward the positive island during its cancellation course. Its
distance from the initial site and apparent horizontal velocity at
different time are presented in Fig. 7. The average horizontal
velocity is 1.0 km s$^{-1}$. Although the V/I magnetograms have
been coaligned with the cross-correlation method, there still
exists an uncertainty in determining the magnetic island position.
A position error (one pixel) introduces an error of the velocity
of (one pixel)/(time interval). In order to reduce the velocity
error, we measure the island position every 6 minutes. As the size
of one pixel is 0{\arcsec}.16 and the time interval is 6 minutes,
the error becomes 0.3 km s$^{-1}$.

Hinode$/$SP also observed the cancellation regions. The vector
magnetic fields and Doppler velocities derived from the SP data
help us to investigate the physical essence of magnetic field
evolution at the cancellation sites. Figure 8 shows the appearance
of three cancellation stages in small sub-regions observed with
SP. From left to right: longitudinal fields
($\alpha_{mag}Bcos\gamma$), transverse fields
($(\alpha_{mag})^{1/2}Bsin\gamma$), inclinations ($\gamma$) and
Doppler velocities ($V_{los}$). Top panels show the properties of
four parameters at the cancellation area at 11:24 UT 04 March. In
the longitudinal magnetogram (top left panel), the cancellation
takes place at the area between the positive and negative elements
(outlined by the parallelogram), and the transverse fields are
strong and point directly from the positive island to the negative
one (second panel). The third panel shows magnetic field
inclinations where the black areas (inclination of 90$\degr$)
indicate horizontal orientations and white areas (0$\degr$ and
180$\degr$) vertical ones. At the polarity inversion line (dotted
curve), the magnetic fields are nearly horizontal. Within the
parallelogram region in the Dopplergram (top right panel), larger
Doppler red-shifts with a mean downward velocity of 1.15 km
s$^{-1}$ are observed. The second and third rows exhibit other two
cancellation stages similar to the first one, and the average
Doppler velocities within the parallelogram regions are 0.84 km
s$^{-1}$ and 0.70 km s$^{-1}$, respectively.

\section{Conclusions and discussion}

Using coordinated SOHO, STEREO and Hinode observations, we
investigate the evolution of two dipoles in a CH region. The
negative element of the first dipole disappears due to its
interaction with the positive element of the second dipole. During
this process, a jet and a plasma eruption are observed. Two
opposite poles of the second dipole constantly emerge and separate
at first, and then cancel with each other. With the decrease of
magnetic flux caused by cancellation, the brightness in 171 {\AA}
images decreases and coronal loops shrink obviously. At the
cancellation sites of the second dipole, the transverse fields are
strong and point directly from the positive elements to the
negative ones. Larger Doppler red-shifts are also observed between
the cancelling elements. At last, the northeastern part of the CH,
where the dipoles are located, appears as a quiet region.

Based on the observational results in this study, we consider that
the interaction between the two dipoles is caused by flux
reconnection, while that between the opposite polarities of the
second dipole is due to the submergence of original loops. This
phenomenon is first reported in a CH. To illustrate the evolution
process of the two dipoles, a series of cartoons are sketched out
(see Fig. 9). Dipole ``1" appears first and dipole ``2" emerges
later at the adjacent area of dipole ``1" (Fig. 9a). We mark the
positive and negative elements of dipole ``1" (``2") with ``A" and
``B" (``C" and ``D"), respectively. Magnetic flux reconnection
occurs between two groups of loops connecting the opposite
polarities. Magnetic loops are restructured to a configuration of
lower potential energy accompanied with energy release. Meanwhile,
small loops form and submerge, leading to an observational
phenomenon, magnetic flux cancellation. Element ``B" totally
disappears due to the cancellation with part of element ``C". Two
poles ``C" and ``D", which are originally connected by flux loops,
draw back and cancel with each other due to flux submergence (Fig.
9c).

Observational evidences, e.g. the jet and the plasma eruption
exhibited in Fig. 3 indicate that the interaction between elements
``B" and ``C" represents a flux reconnection process. When
magnetic reconnection occurs, magnetic energy is converted into
thermal energy and kinetic energy. Then plasma jet forms and
ejects along the field line (Yokoyama \& Shibata 1995).
Accompanying the reconnection, plasma eruption may also be formed
due to the change of magnetic configuration. However, we can not
absolutely rule out the possibility that the disappearance of
element ``B" is caused by the emergence of U$-$shaped magnetic
loops connecting ``B" and ``C" from below the photosphere (Parker
1984; Lites et al. 1995), for we lack more relevant observational
information.

Flux loops are always affected by magnetic buoyancy and tension of
subsurface field lines. When the tension gains the upper hand, the
flux loops are pulled back down by magnetic tension and submerge
(Zirin 1985). The movies of 1-min cadence MDI magnetograms and
3-min cadence Na V/I magnetograms and the results revealed in
Figs. 4$-$8 let us believe that the cancellation between ``C" and
``D" is caused by the submergence of original loops. From Fig. 7,
we obtain an average horizontal velocity of 1.0 km s$^{-1}$. When
we assume flux loops connecting the cancelling elements are
approximately semicircular, the vertical velocity of submergence
is 1.0 km s$^{-1}$. At the cancellation sites in this study, we
observe a mean downward velocity of 0.9 km s$^{-1}$ (averaging the
values of 1.15 km s$^{-1}$, 0.84 km s$^{-1}$ and 0.70 km s$^{-1}$
in the three stages), much higher than that of the surrounding
areas (Fig. 8), comparable to the velocity obtained from the
cancelling flux motion in Fig. 7 (1.0 km s$^{-1}$) and consistent
with the velocities reported by Harvey et al. (1999) (0.3$-$1.0 km
s$^{-1}$) and Chae et al. (2004) (1.0 km s$^{-1}$).

Evolution of CHs refers to several aspects, such as CH formation
and decay, and the temporal variation of CH boundary. The key
quantity for understanding these aspects is magnetic field.
Magnetic flux emergence, cancellation, mergence and dispersion are
the main forms of magnetic field evolution. They are all found in
the dipolar evolution process in the CH in this study. An
interesting cancelling form, i.e. submergence of initial loops
after emergence, is also observed for the first time in the CH. At
the late stage of the dipolar evolution, the area where the
dipoles are located becomes mixed polarities instead of unipolar
fields, resulting in the change of the overlying corona from a CH
area to a quiet region (see the rectangle region in Fig. 1). This
confirms the result of Zhang et al. (2007) that one of the
signatures of decay of a CH is the disappearance of the magnetic
flux imbalance. These results enlighten us that, in order to
understand the CH evolution, it is important to study magnetic
field evolution in CHs. In particular, to investigate the
evolution of dipoles may be an efficient approach to understand CH
decay and disappearance.

\acknowledgments The authors are grateful to the anonymous referee
for the constructive comments and detailed suggestions to improve
this manuscript. We acknowledge the {\it SOHO}, {\it STEREO} and
{\it Hinode} teams for providing the data. {\it SOHO} is a project
of international co-operation between ESA and NASA, and {\it
STEREO} a NASA project. {\it Hinode} is a Japanese mission
developed and launched by ISAS/JAXA, with NAOJ as domestic partner
and NASA and STFC (UK) as international partners. It is operated
by these agencies in co-operation with ESA and NSC (Norway). This
work is supported by the National Natural Science Foundations of
China (G40674081, 40890161, 10703007, and 10733020), the CAS
Project KJCX2-YW-T04, the National Basic Research Program of China
under grant G2006CB806303, and the Young Researcher Grant of
National Astronomical Observatories, Chinese Academy of Sciences.

\clearpage

\begin{table}
\begin{center}
\caption{Data sets used in this study.\label{tbl-1}}
\centering
\begin{tabular}{ccccc}
\tableline\tableline
Observation & Period & Cadence & Pixel Size & Field of View \\
 & (UT) & (minutes) & (arcsec) & (arcsec$^{2}$) \\
\tableline
SOHO$/$MDI & 02 08:03 -- 07 12:47 & 96 & 1.97 & full disk \\
 & 02 11:35 -- 02 14:36 & 1 & 1.97 & full disk \\
 & 03 00:57 -- 07 12:00 & 1 & 1.97 & full disk \\
SOHO$/$EIT(195{\AA}) & 02 08:00 -- 07 12:00 & 12 & 2.63 & full disk \\
STEREO-B$/$EUVI(171{\AA}) & 02 08:00 -- 05 00:00 & 2.5 & 1.59 & full disk \\
STEREO-A$/$EUVI(171{\AA}) & 05 00:00 -- 07 12:00 & 2.5 & 1.59 & full disk \\
Hinode$/$SP(Fe full Stokes) & 03 12:00 -- 14:42 & 50\tablenotemark{a} & 0.32 & 59.04${\times}$162.30 \\
 & 04 11:20 -- 15:55 & 50\tablenotemark{a} & 0.32 & 59.04${\times}$162.30 \\
 & 05 11:40 -- 15:15 & 45\tablenotemark{a} & 0.32 & 59.04${\times}$162.30 \\
Hinode$/$NFI(Na Stokes I, V) & 03 10:53 -- 15:27 & 3 & 0.16 & 163.84${\times}$163.84 \\
 & 04 11:20 -- 16:17 & 3 & 0.16 & 64.00${\times}$163.84 \\
 & 05 11:39 -- 15:27 & 3 & 0.16 & 64.00${\times}$163.84 \\

 \tableline
\end{tabular}
\tablenotetext{a}{Scan time for one SP map is 12.5 minutes.}
\end{center}
\end{table}

\clearpage

\begin{figure}
\centering
\includegraphics[bb=27 240
575 588,clip,angle=0,scale=0.85]{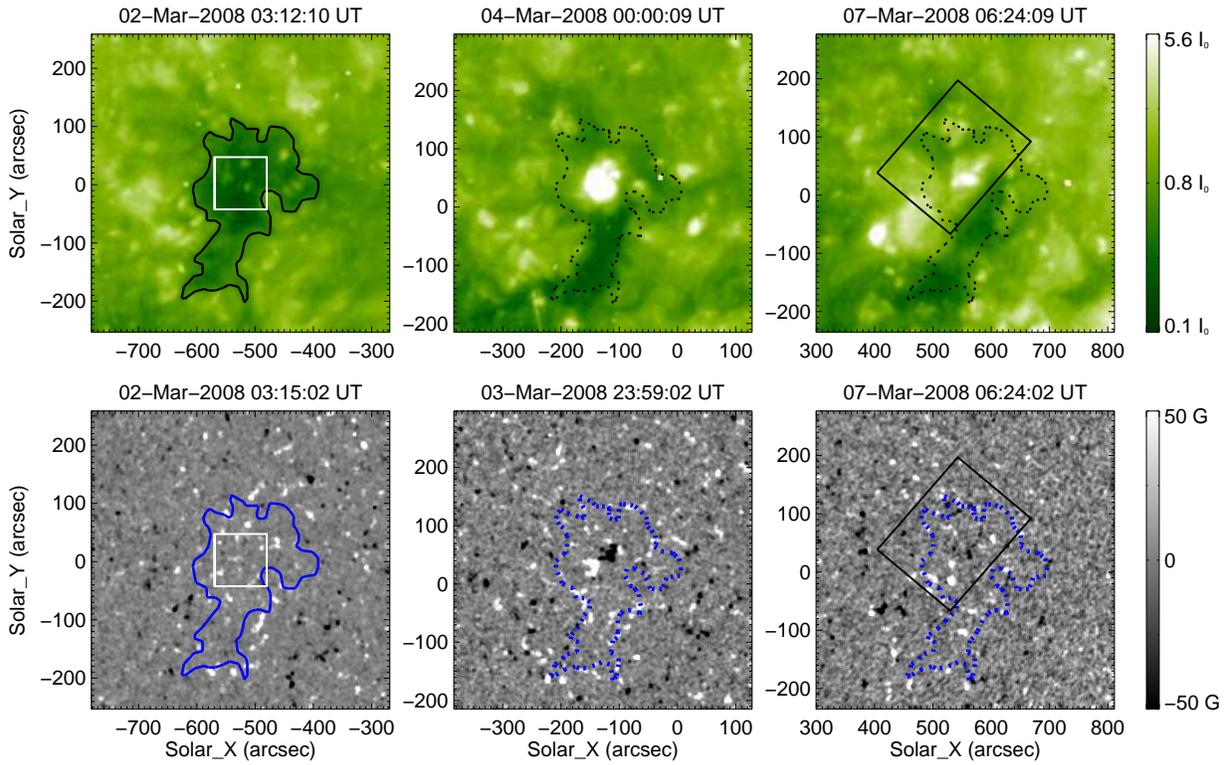} \caption{EIT 195 {\AA}
images (\emph{top panels}) and MDI magnetograms (\emph{bottom
panels}) showing the evolution of the CH. Black and blue curves
delineate the CH boundary derived from the EUVI 284 {\AA} image
obtained at 03:06:30 UT 02 March 2008. The squares outline the
field-of-view of Figs. 2$-$4 and rectangles enclose a QS region.
\label{fig1}}
\end{figure}
\clearpage

\begin{figure}
\centering
\includegraphics[bb=42 260
572 571,clip,angle=0,scale=0.85]{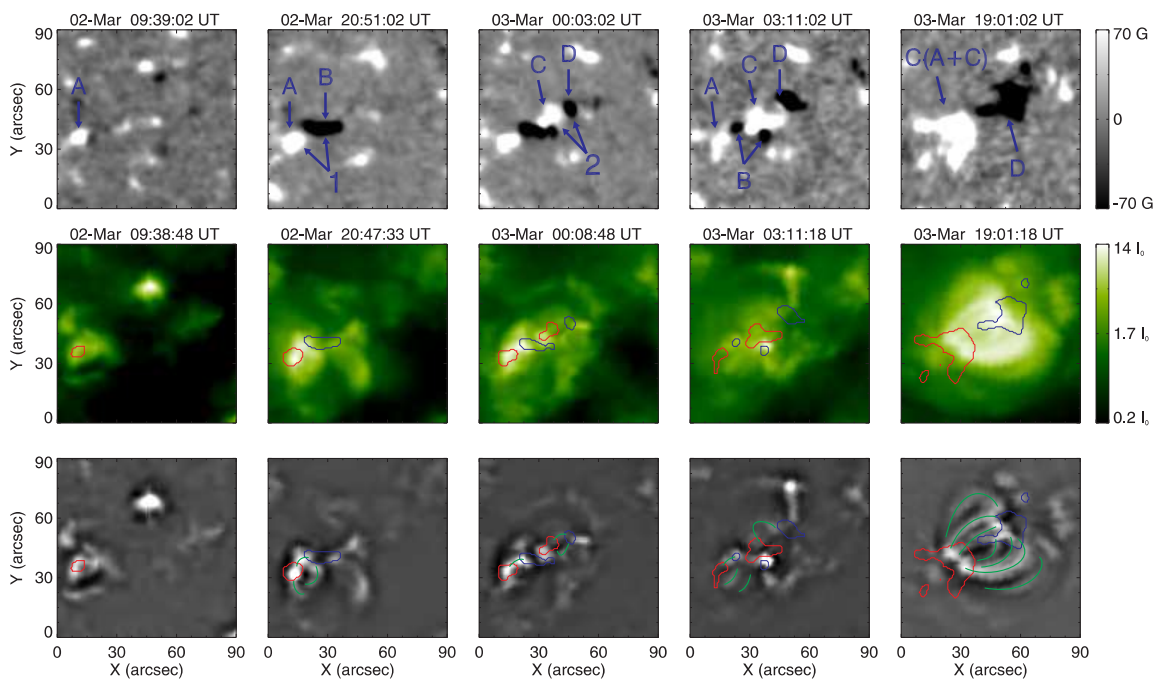} \caption{Interaction
between two dipoles. \textit{From top to bottom}: MDI
magnetograms, EUVI 171 {\AA} images and corresponding contrast
enhanced images. Red and blue curves are contours of the positive
($+$70 G) and negative ($-$70 G) magnetic fields, while green
curves emphasize and figure out loop connections. Arrows ``1" and
``2" denote two dipoles emerging one after the other. We denote
the positive and negative elements of dipole ``1" (``2") with
arrows ``A" and ``B" (``C" and ``D"), respectively. \label{fig2}}
\end{figure}
\clearpage

\begin{figure}
\centering
\includegraphics[clip,angle=0,scale=1.0]{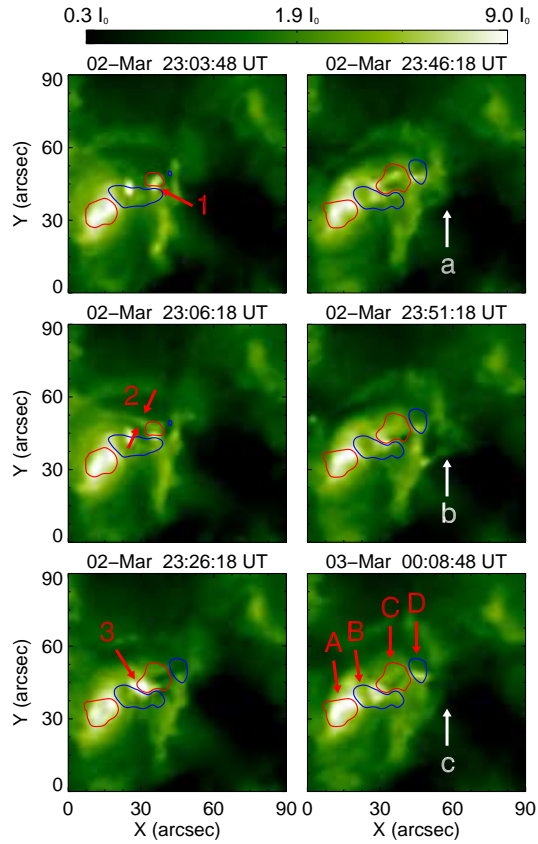} \caption{EUVI
images displaying a jet and a plasma eruption. ``A"--``D" are
contours of two dipoles at $\pm$30 G levels from MDI magnetogram
at 22:27 UT 02 March (first two panels) and at 00:03 UT 03 March
(other panels). Arrows ``1" and ``3" denote two brightening
points, while arrows ``2" indicate an EUV jet. Arrows ``a"--``c"
show the different stages of a plasma eruption. \label{fig3}}
\end{figure}
\clearpage

\begin{figure}
\centering
\includegraphics[bb=45 246
560 610,clip,angle=0,scale=0.85]{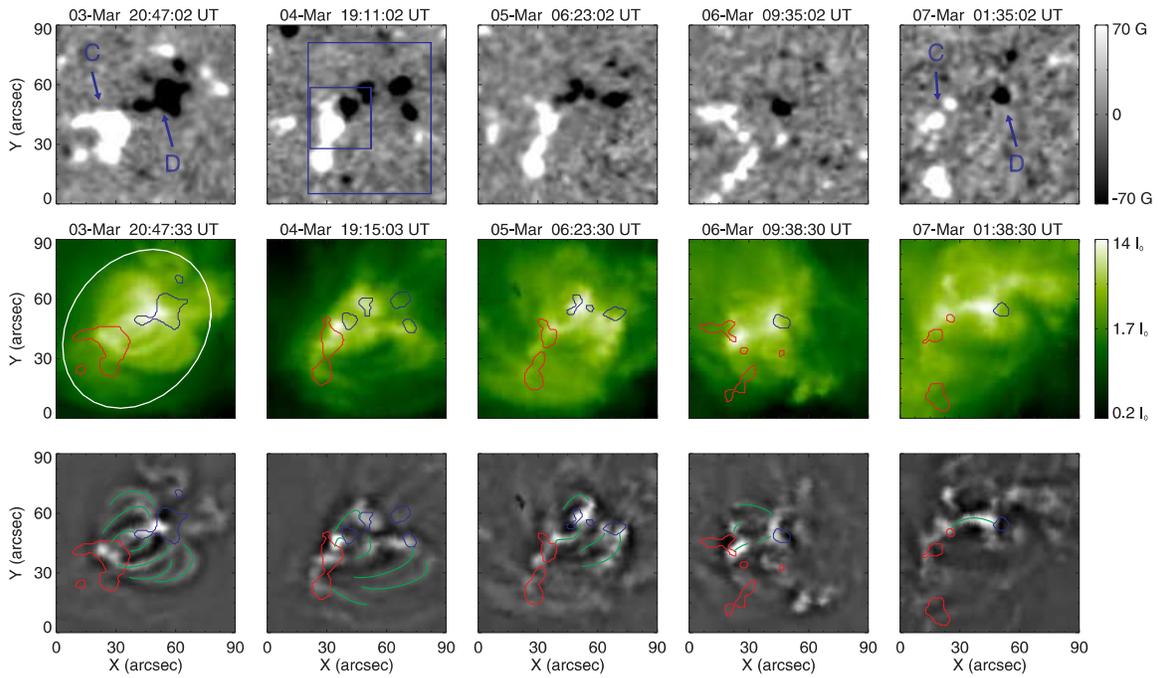} \caption{Similar to Fig.
2 but for magnetic flux cancellation between two polarities of
dipole ``2". Small square on the second MDI magnetogram outlines
the field-of-view of Fig. 6 and larger rectangle the area where
the positive and negative flux of the second dipole are measured
in the Na V/I magnetograms. Ellipse outlines the region where
magnetic flux in MDI magnetograms and brightness in 171 {\AA}
images are measured.\label{fig4}}
\end{figure}
\clearpage

\begin{figure}
\centering
\includegraphics[bb=91 269
495 553,clip,angle=0,scale=1.0]{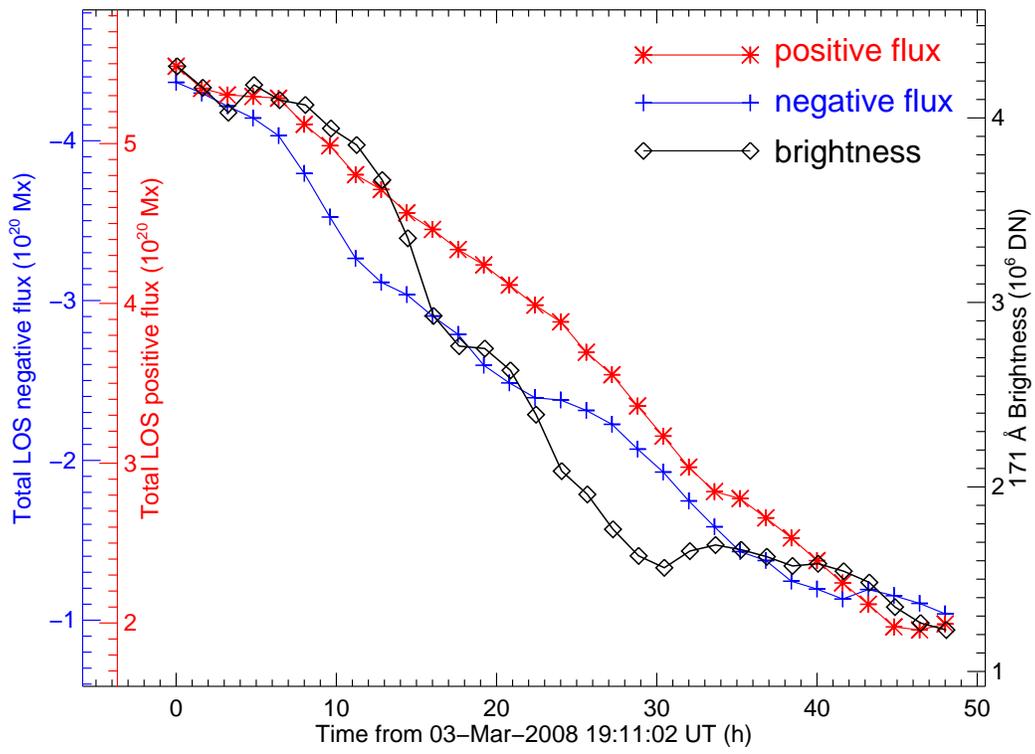} \caption{Temporal
variations of negative and positive magnetic flux in MDI
magentograms and brightness in 171 {\AA} images derived from the
ellipse region in Fig. 4. \label{fig5}}
\end{figure}
\clearpage

\begin{figure}
\centering
\includegraphics[bb=31 234
580 631,clip,angle=0,scale=0.8]{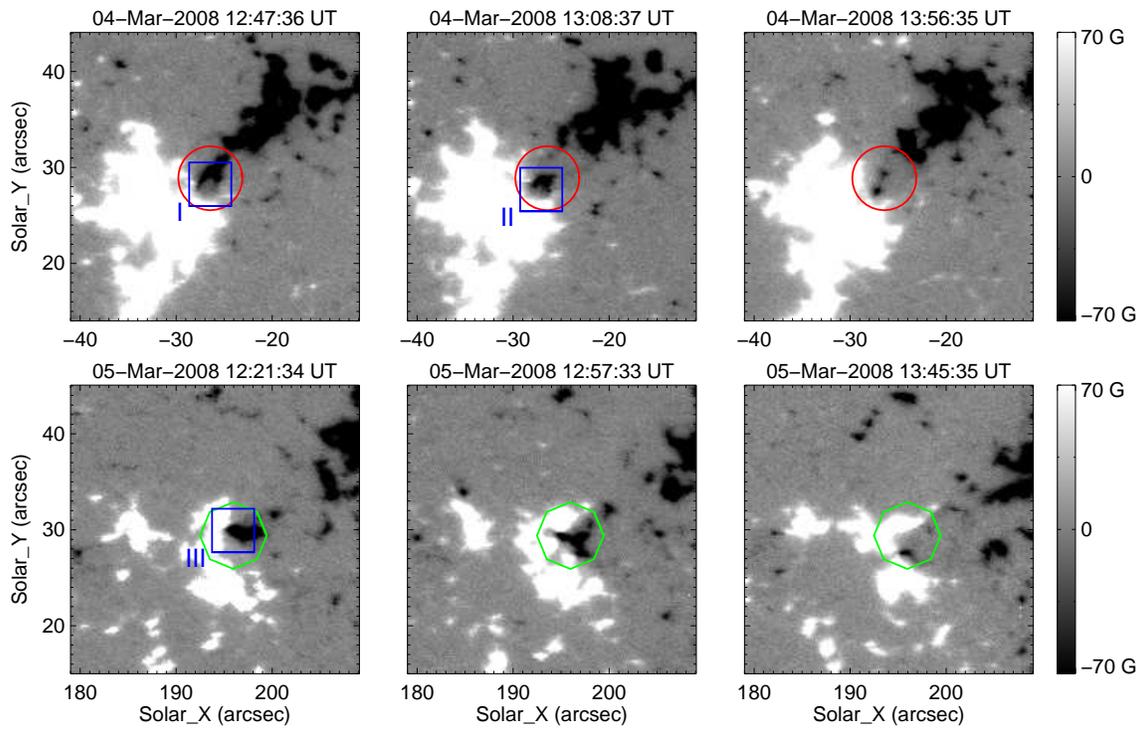} \caption{Time sequence of
Hinode$/$NFI Na V/I magnetograms displaying magnetic flux
cancellation. Circles and octagons show two cancellation events on
two days, respectively. Rectangles ``I" ``II" and ``III" outline
the locations of SP maps from top to bottom in Fig. 8.
\label{fig6}}
\end{figure}
\clearpage

\begin{figure}
\centering
\includegraphics[bb=75 248
515 573,clip,angle=0,scale=1.0]{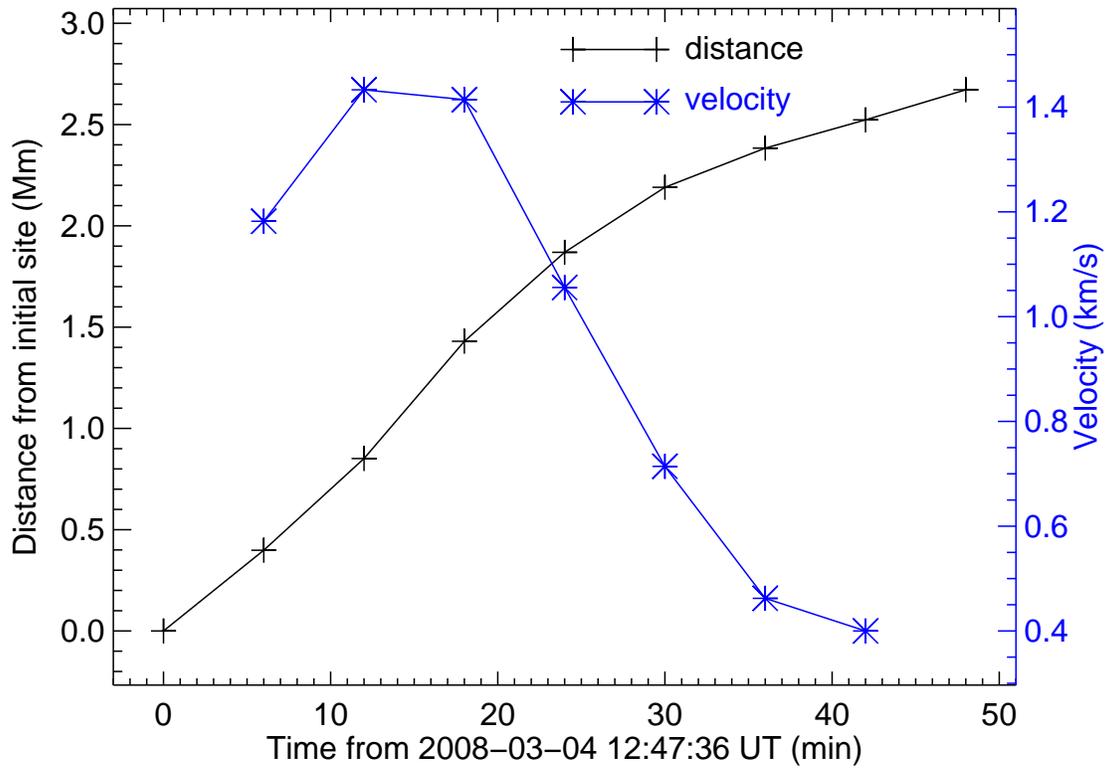} \caption{Temporal
variations of distance and apparent horizontal velocity of the
negative island in the circle region shown in Fig. 6.
\label{fig7}}
\end{figure}
\clearpage

\begin{figure}
\centering
\includegraphics[bb=39 214
550 650,clip,angle=0,scale=0.92]{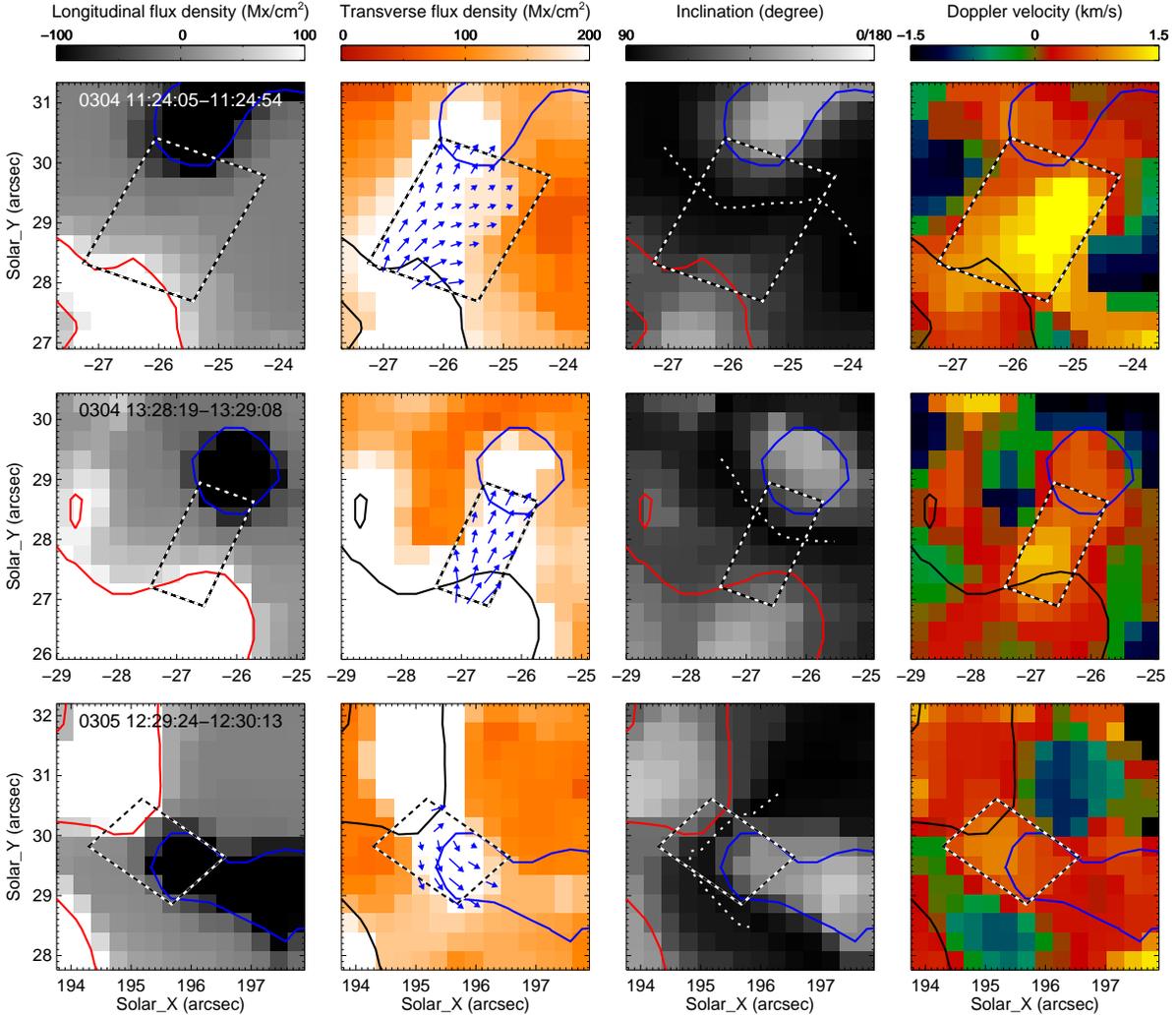} \caption{Appearance of
cancellation regions observed by Hinode$/$SP. \textit{From left to
right}: longitudinal fields, transverse fields, inclinations and
Doppler velocities. Inclinations of 90$\degr$ correspond to
magnetic fields with horizontal orientations, and positive Doppler
velocities to red-shifts. Blue curves are contours of negative
elements ($-$100 G) and other solid curves outline the positive
ones (+100 G). Dotted curves represent the polarity inversion
lines. Arrows denote the transverse fields and parallelograms
outline the areas where cancellations take place. \label{fig8}}
\end{figure}
\clearpage

\begin{figure}
\centering
\includegraphics[scale=1.0]{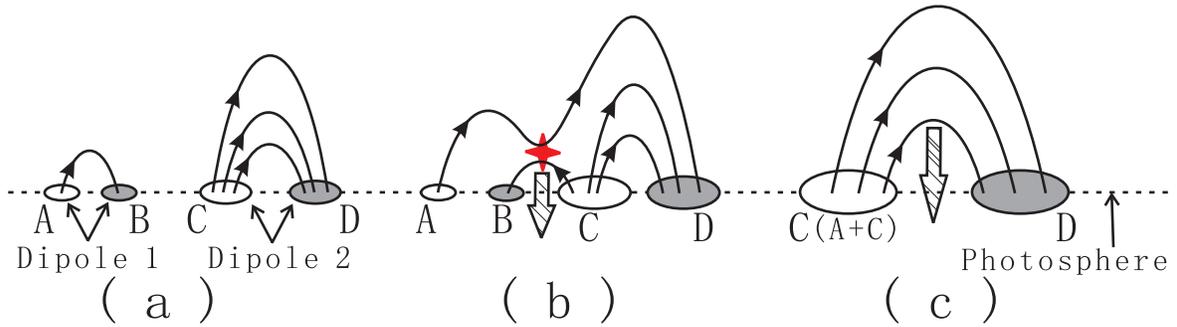} \caption{Cartoons illustrating
evolution process of the two dipoles. (a) Pre-interaction state of
the dipoles. (b) Flux disappearance due to reconnection
accompanied with energy release. (c) Flux cancellation caused by
submergence of original loops connecting the dipolar elements.
``A" and ``B" (``C" and ``D") represent the positive and negative
elements of dipole ``1" (``2"), respectively. Asterisk marks
magnetic flux reconnection and cross hatched arrows indicate flux
submergence.\label{fig9}}
\end{figure}
\clearpage

\end{document}